\documentclass[preprint,aps,floatfix]{revtex4}
\usepackage{graphicx,epsfig,subfig,dcolumn,bm,mathrsfs,amsmath}
\newcommand{\ud}{\mathrm{d}}

\begin{document}

\title{Shape variation of micelles in polymer thin films}    

\author{Jiajia Zhou}
\altaffiliation{Present address: Komet331, Institut f\"ur Physik, Johannes Gutenberg-Universit\"at Mainz, Staudingerweg 7, D55099 Mainz, Germany}
\email[]{zhou@uni-mainz.de} 
\affiliation{Department of Physics \& Astronomy, McMaster University \\
Hamilton, Ontario, Canada L8S 4M1}

\author{An-Chang Shi}
\email[]{shi@mcmaster.ca}
\affiliation{Department of Physics \& Astronomy, McMaster University \\
Hamilton, Ontario, Canada L8S 4M1}


\begin{abstract}
The equilibrium properties of block copolymer micelles confined in polymer thin films are investigated using self-consistent field theory.
The theory is based on a model system consisting of AB diblock copolymers and A homopolymers. 
Two different methods, based on the radius of gyration tensor and the spherical harmonics expansion, are used to characterize the micellar shape. 
The results reveal that the morphology of micelles in thin films depends on the thickness of the thin films and the selectivity of the confining surfaces. 
For spherical (cylindrical) micelles, the spherical (cylindrical) symmetry is broken by the presence of the one-dimensional confinement, whereas the top-down symmetry is broken by the selectivity of the confining surfaces. 
Morphological transitions from spherical or cylindrical micelles to cylinders or lamella are predicted when the film thickness approaches the micellar size.
\end{abstract}


\maketitle

\section{introduction}
\label{sec:introduction}

Polymer thin films are widely used in technology and industry.
Applications of polymer thin films include protective surface coatings, birefringent elements in optical devices, and printing circuits in flexible solar cells.  
Films made by blending different polymers provide an economical route to obtain new products with enhanced properties. 
In general, application of polymer films requires the surface properties to be homogeneous, but polymer blends tend to phase separate.  
To improve the dispersion of the polymer blends, compatibilizers such as diblock copolymers are commonly used to induce homogeneous mixing between immiscible homopolymers. 
However, block copolymers tend to form local aggregates, or micelles, when the copolymer concentration exceeds a critical micelle concentration. 
The micelle formation is driven by the unequal miscibility of the two blocks with the homopolymers, such that the miscible blocks form a corona whereas the immiscible blocks form a core, thus minimizing the unfavorable interactions. 
The process of micelle formation is similar to the segregation of surfactants in water and the formation of bilayers from amphiphilic lipids.  
Depending on the chain architecture and interaction parameters, copolymer can from micelles of different shapes. 
Common morphologies include bilayer sheets, cylinders or spheres \cite{Safran}.  

In some processes, micelle formation is an undesirable effect which should be avoided, but there are also applications where polymer micelles can beneficial. 
One example is to use micelles for drug delivery by encapsulating the medicine substance inside the micelle core \cite{Kim2005}. 
Compared to micelles made by surfactants of low-molecular weight, polymeric micelles are significantly more stable, have a larger capacity, and can be biocompatible by modifying the chemical details.    
Under biological conditions, micelles need to pass through various confined environments, such as the capillary vessels. 
Therefore, it is crucially important to understand how confinement affects the micelle properties.

Polymer thin films can be viewed as polymers under planar or one-dimensional confinement. 
For copolymer melts, planar confinement can introduce frustration in the bulk structure and induce novel morphologies which are not available in bulk system \cite{Lambooy1994, Kellogg1996, Harrison1998, Radzilowski1998, Knoll2002}. 
The phase behavior is determined by the relation between the size of confinement and the natural spacing of the structure. 
When the two length scales are compatible, bulk structures are preserved; otherwise, new morphologies such as perforated lamellae can be formed. 
Similarly, the confinement can also influence the micelle formation in diblock copolymer/homopolymer blends, and the important factor is the ratio between the micelle size and the dimension of confinement. 
When the confinement is weak, the micelle reacts by slightly shrinking its size in the direction of the confinement. 
As the dimension of the confinement becomes compatible to the micelle size, the micelle formation becomes unfavorable, resulting in an increase of the critical micelle concentration \cite{Zhu1999,ZhangXianren2007}.   

Various computational methods have been used to study diblock copolymers under planar confinement, such as the scaling argument \cite{Turner1992, Walton1994}, simulated annealing \cite{YinYuhua2004, YinYuhua2006}, Monte Carlo simulations \cite{WangQiang2000, WangQiang2001}, density function theory \cite{Huinink2001} and self-consistent field theory \cite{Shull1992a, ChenPeng2007, MengDong2010, ManXingkun2013, LiWeihua2013}.
Most of these studies have been focused on different morphologies of the copolymer melts or copolymer blends, but specific studies of the micelle formation in polymer thin film have been limited. 
Zhu \emph{et. al.} used a lattice-base self-consistent field theory to study the confinement-induced miscibility in polymer blends \cite{Zhu1999}. 
They focused on a spherical micelle consisted of symmetric diblock copolymers and found the critical micelle concentration increases dramatically when the surface separation is smaller than the micelle size. 
Similar trends were observed by Zhang \emph{et. al.} using lattice Monte Carlo simulations \cite{ZhangXianren2007}, and they also investigated the influence of the surface interaction on critical micelle concentration.
Cavallo \emph{et. al.} employed Monte Carlo simulations of the bond fluctuation model in a confined blends of asymmetric diblock copolymer and homopolymer \cite{Cavallo2008}. 
They have considered the effect of the surface interaction and copolymer concentration. 
These studies have greatly enriched our understanding of the micelle formation in the thin film geometry. 
However, a detailed investigation of the shape variation by changing various parameters, especially the surface separation, has been lacking.
Furthermore the effect of the asymmetric surfaces is remained to be explored. 

In order to better understand the micelle formation and shape variation in thin film geometry, we employ a real-space self-consistent field theory to study the micelle shape in a confined blend of diblock copolymers and homopolymers (AB/A).  
Micelles have different morphologies in a bulk system: bilayers, cylinders, and spheres \cite{2011_cmc_bulk}.   
Only the bilayer structure is commensurate with the thin film geometry, whereas the spherical and cylindrical micelles must undergo deformation when perturbed by the surface. 
For simplicity, we choose to focus on two cases in this work: one is a spherical micelle, and the other one is a cylindrical micelle with its axis parallel to the surface. 
Our paper is organized as follows: In the next section, we introduce the simulation model and system parameters. 
The deformation of the micelle is discussed in Section \ref{sec:results}, and the effects of various parameters are elucidated. 
We summarize our results in Section \ref{sec:summary}.

\section{Self-consistent Field Theory}
\label{sec:scft}

Self-consistent field theory (SCFT) is a versatile and accurate method to study the phase behavior of polymeric systems. 
The theory has been well described in several excellent reviews and monographs \cite{Schmid1998, Matsen2002, Shi2004_chapter, Fredrickson}, and we refer readers to these references for details.
In this section, we describe briefly the implementation of SCFT in a confined geometry and introduce the notation and controlling parameters.

The polymer blend is composed of diblock copolymers AB and homopolymers A in a volume of $V$.  
We assume the copolymer and homopolymer chains have the same length $N$, and the A-monomer (B-monomer) fraction of the copolymer is $f_A$ ($f_B=1-f_A$). 
The blend is incompressible, and all monomers have the same volume $\rho_0^{-1}$ and Kuhn length $b$. 
In the following, lengths will be expressed using the unit $\sqrt{N}b$ and the energies are scaled by $k_BT$.
The interaction between A/B monomers is characterized by the standard Flory-Huggins parameter $\chi$, and an intermediate segregation case is considered at $\chi N=20$. 
We formulate our theory in the grand canonical ensemble and use the chemical potential of homopolymers as a reference state. 
Therefore the controlling parameter is the copolymer chemical potential $\mu_c$, or its activity $z_c=\exp(\mu_c)$.
We also describe the blend composition using the bulk copolymer concentration $\phi_c^{\rm bulk}$, which is related to $\mu_c$ by
\begin{equation}
   \mu_c = \ln \frac{ \phi_c^{\rm bulk} } {1-\phi_c^{\rm bulk}} + f_B \chi N (1-2f_B \phi_c^{\rm bulk}).   
\end{equation}

Within the SCFT framework, the interactions between many chains are replaced by the interactions between one ideal Gaussian chains and an effective mean-field potential.
The grand free energy can be written as a functional of local densities $\{\phi\}$ and their conjugate fields $\{\omega\}$,  
\begin{eqnarray}
 \label{eq:freeE}
  \frac{N\mathscr{F}}{k_BT\rho_0} &=& \int \ud \mathbf{r} \,
  [\chi N \phi_{A}(\mathbf{r}) \phi_{B}(\mathbf{r})] \\
  && - \int \ud \mathbf{r} \, [ \omega_{A}(\mathbf{r})
  \phi_{A}(\mathbf{r}) + \omega_{B}(\mathbf{r}) \phi_{B}(\mathbf{r})]
  \nonumber \\ 
  && - \int \ud \mathbf{r} \, H(\mathbf{r}) [ \phi_A(\mathbf{r}) -
  \phi_B(\mathbf{r}) ] \nonumber \\
  && - \int \ud \mathbf{r} \, \xi(\mathbf{r})
  [\phi_0(\mathbf{r}) -\phi_A(\mathbf{r})-\phi_B(\mathbf{r})] \nonumber \\
  && - z_c Q_c - Q_h. \nonumber
\end{eqnarray}
The first line in the right-hand side is the intermolecular interactions.
The second line presents the coupling between the local densities and their conjugate fields.
The third line is the contribution due to the surface interaction, where $H(\mathbf{r})$ is the external surface field.
The fourth line introduces a Lagrange multiplier $\xi(\mathbf{r})$ to enforce the local total density to be $\phi_0(\mathbf{r})$. 
The exact forms of $H(\mathbf{r})$ and $\phi_0(\mathbf{r})$ will be discussed late in this section.
The two terms in the last line are the configuration entropies, related to the single-chain partition functions for the copolymer ($Q_c$) and the homopolymer ($Q_h$).   
For the copolymer, the partition function has a form $Q_c = \int \ud
\mathbf{r} q_c(\mathbf{r},1)$, where $q_c(\mathbf{r},s)$ is an end-integrated propagator, and $s$ is the normalized arc-length runs from 0 to 1.  
The propagator satisfies the modified diffusion equation
\begin{equation}
  \label{eq:diffusion}
  \frac{\partial}{\partial s} q_c(\mathbf{r},s) = 
  \frac{1}{6} N b^2 \nabla^2 q_c(\mathbf{r},s) 
  - \omega(\mathbf{r}) q_c(\mathbf{r},s) ,
\end{equation}
where the field is given by
\begin{equation}
  \omega(\mathbf{r}) = 
  \begin{cases}
    \omega_A(\mathbf{r}) & \text{if } 0<s<f_A, \\
    \omega_B(\mathbf{r}) & \text{if } f_A<s<1. 
  \end{cases}
\end{equation}
The initial condition is $q_c(\mathbf{r},0)=1$. 
Since the copolymer has two distinct ends, a complementary end-integrated
propagator $q_c^+(\mathbf{r},s)$ is introduced. 
It satisfies Eq.~(\ref{eq:diffusion}) with the right-hand side multiplied by $-1$, and the initial condition $q_c^+(\mathbf{r},1)=1$. 
For the homopolymer, one propagator $q_h(\mathbf{r},s)$ is sufficient, and the single-chain partition function has a form $Q_h = \int \ud \mathbf{r}\, q_h(\mathbf{r},1)$.

To proceed, SCFT employs a mean-field approximation, which amounts to evaluate the free energy using a saddle-point technique. 
Technically the saddle-point approximation is obtained by demanding that the functional derivatives of expression (\ref{eq:freeE}) to be zero,
\begin{equation}
  \frac {\delta \mathscr{F}} {\delta \phi_{\alpha}}  
  = \frac {\delta \mathscr{F}} {\delta \omega_{\alpha}}  
  = \frac {\delta \mathscr{F}} {\delta \xi} =0.  
\end{equation}
These conditions lead to the following mean-field equations,
\begin{eqnarray}
  \phi_{A} (\mathbf{r}) &=& \int_0^1 \ud s\, q_{h}
  (\mathbf{r},s) q_{h} (\mathbf{r}, 1-s) \\
  && + z_c \int_{0}^{f_A} \ud s\,
  q_{c} (\mathbf{r},s) q_{c}^+ (\mathbf{r}, s), \nonumber\\ 
  \phi_{B} (\mathbf{r}) &=& z_c \int_{f_A}^{1} \ud s\, q_{c} 
  (\mathbf{r},s) q_{c}^+ (\mathbf{r},s),\\
  \omega_A (\mathbf{r}) &=& \chi N \phi_B (\mathbf{r}) 
  - H(\mathbf{r}) + \xi (\mathbf{r}), \\ 
  \omega_B (\mathbf{r}) &=& \chi N \phi_A (\mathbf{r}) 
  + H(\mathbf{r}) + \xi (\mathbf{r}), \\ 
  \phi_0(\mathbf{r}) &=& \phi_A(\mathbf{r}) + \phi_B(\mathbf{r}).
\end{eqnarray}
These equations can be solved using iteration. 
The resulting solutions then can be substituted into Eq.~(\ref{eq:freeE}) to compute the free energy.
We are interested in the free energy of a system containing one micelle $(\mathscr{F})$ compared to that of a reference system without the micelle $(\mathscr{F}')$.
Therefore we define an excess free energy, $F_{\rm micelle}$, as the free energy difference,
\begin{equation}
  \label{eq:freeE_micelle}
  F_{\rm micelle} = \frac{ N(\mathscr{F} - \mathscr{F}') }{ k_BT \rho_0 }.
\end{equation}
Note that we have scaled the free energy in terms of thermal energy $k_BT$ and number density of the molecules $\rho_0/N$.

\begin{figure}[htp]
  \centering
  \includegraphics[width=0.80\columnwidth]{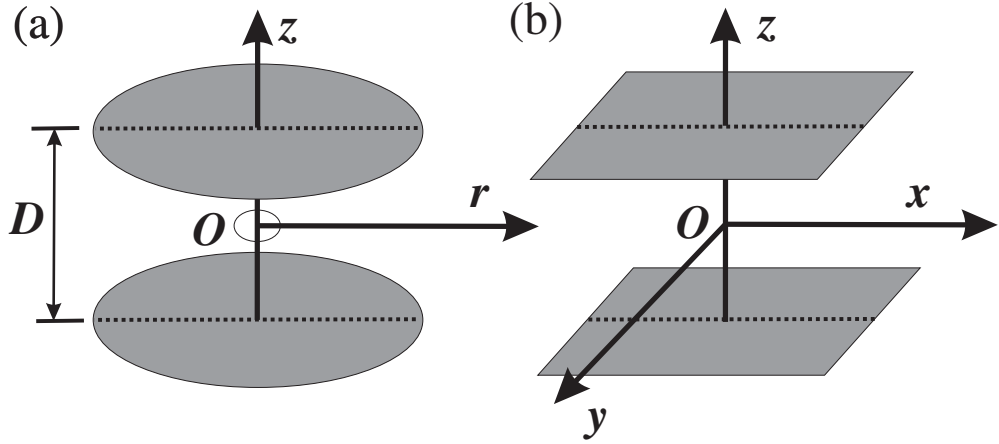}
  \caption{Schematics of the coordinate systems. (a) A cylindrical coordinate system $Orz$ used for micelles with a rotational symmetry, i.e. a sphere or a perpendicular cylinder. (b) A Cartesian coordinate system $Oxz$ for micelles whose properties do not vary in $y$-direction, such as a cylinder parallel to the surface and a bilayer.}
  \label{fig:coordinates}
\end{figure}

We consider two micelle morphologies . 
The first case is the spherical micelle formed in a blend with $f_A=0.7$ and $\phi_c^{\rm bulk}=0.11$. 
The parameters are chosen such that spherical micelle is the stable morphology in an unconfined system \cite{2011_cmc_bulk}. 
To model the spherical micelle under planar confinement, we adapt a cylindrical coordinate system $\mathbf{r}=(r, z)$, which has the rotational symmetry with respect to $z$-axis.  
The two hard surfaces are located at $z=-D/2$ and $z=D/2$.
Figure~\ref{fig:coordinates}(a) shows a schematic setup of the cylindrical coordinate system. 
In the second case, we consider a cylindrical micelle whose axis lies parallel to the hard surfaces. 
The polymer blend has a composition of $\phi_c^{\rm bulk}=0.016$ and $f_A=0.6$. 
A two-dimensional Cartesian coordinates $\mathbf{r}=(x,z)$ is used.
Figure~\ref{fig:coordinates}(b) shows the coordinate system, where $y$-axis is along the micelle central axis.

\begin{figure}[htp]
   \centering
   \includegraphics[width=0.8\columnwidth]{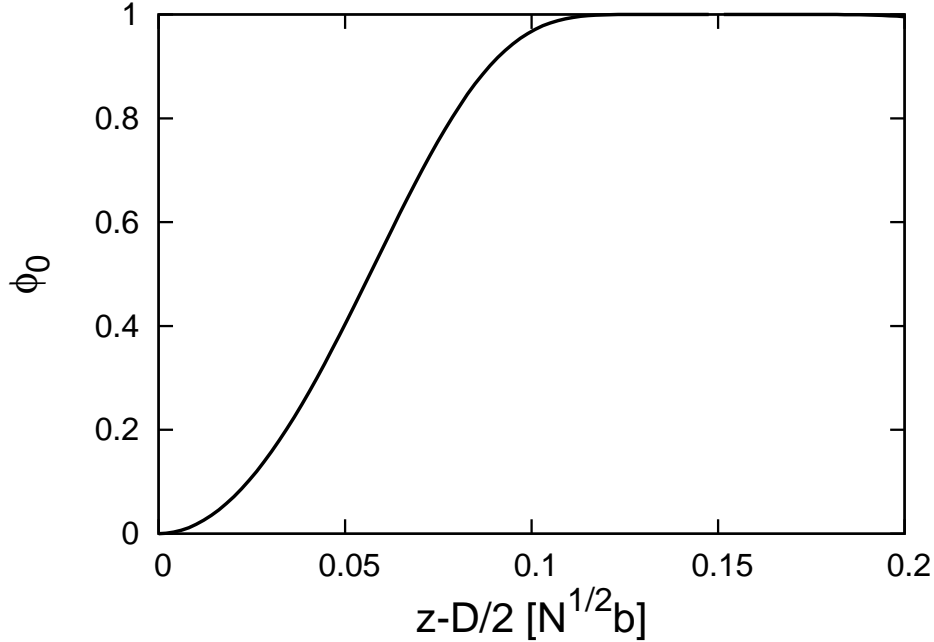}
   \caption{Total density profile near the lower surface. The characteristic length of the surface layer is $\tau=0.15 \sqrt{N}b$.}
   \label{fig:hard_surface}
 \end{figure}

In the general SCFT framework, the incompressibility is enforced by setting the total monomer density $\phi_0(\mathbf{r})=\phi_A(\mathbf{r})+\phi_B(\mathbf{r})=1$ everywhere. 
Under planar confinement, the top and bottom surfaces are impenetrable, and the total density is required to vanish at the surfaces. 
A common method to solve this problem is to use a profile $\phi_0(\mathbf{r})$ that is $1$ everywhere except in the proximity of the surface. 
Close to the surface, the density decreases continuously from $1$ to $0$ over a short distance. 
The detailed form of $\phi_0(\mathbf{r})$ is not important, and we use the one proposed by Meng and Wang \cite{MengDong2007} to enhance the numerical stability,
\begin{equation}
  \phi_0 (\mathbf{r}) = \left\{
  \begin{array}{lrcl} 
    \tanh^2 \left[ \frac{ \textstyle{ 2 \tau (z+\frac{D}{2}) }}
    { \textstyle{ \tau^2 - (z+\frac{D}{2})^2 }} \right] 
    & - \frac{D}{2} \le & z & \le -\frac{D}{2} +\tau \\
    1 & -\frac{D}{2} + \tau \le & z & \le \frac{D}{2}-\tau \\
    \tanh^2 \left[ \frac{ \textstyle{ 2 \tau (\frac{D}{2}-z) }}
    { \textstyle{ \tau^2 - (\frac{D}{2}-z)^2 }} \right]
    & \frac{D}{2}-\tau \le & z & \le \frac{D}{2} 
  \end{array} \right.
\end{equation}
where $\tau$ is the characteristic length of the surface layer.
The total density profile $\phi_0(\mathbf{r})$ near the lower surface is plotted in Fig.~\ref{fig:hard_surface} for $\tau=0.15 \sqrt{N}b$. 

The surface interaction is represented by a surface field $H(\mathbf{r})$. 
For simplicity, a short-range, linear field with the same characteristic length of $\rho_0(\mathbf{r})$ is used:
\begin{equation}
  H(\mathbf{r}) = \left\{
  \begin{array}{lrcl} 
    \Lambda_l \left[ 1- (z+\frac{D}{2})/\tau \right]
    & -\frac{D}{2} \le & z & \le -\frac{D}{2} +\tau \\
    0 & -\frac{D}{2} + \tau \le & z & \le \frac{D}{2} -\tau \\
    \Lambda_u \left[ 1- (\frac{D}{2}-z)/\tau \right] 
    & \frac{D}{2}-\tau \le & z & \le \frac{D}{2} 
  \end{array} \right.
\end{equation}
The constants $\Lambda_l$ and $\Lambda_u$ are used to characterize the
surface selectivity for the lower and upper surfaces, respectively. 
In our notation, a positive value of $\Lambda$ corresponds to the
attractive interaction to A-monomers (or a repulsive interaction to
B-monomers).  

The computation box is chosen large enough such that far away from the micelle, the bulk concentration for copolymers are reached.
In general, $4.0 \sqrt{N}b$ is sufficient in the direction parallel to the surface.  
The separation between two hard surfaces is adjusted from $1.0\sqrt{N}b$ to $8.0\sqrt{N}b$.
The self-consistent equations are solved in the real-space using the alternative direction implicit (ADI) scheme \cite{NR3}, with a grid size of $0.02 \sqrt{N}b$. 
It is also important to set the initial density profile for the start of iterations, and we choose to set the micelle with a tanh profile. 
In addition, we have tested different initial configurations to ensure that the converged solutions do not depend on the initial conditions.

\section{Results and discussion}
\label{sec:results}

In this section, we study the shape variation and morphology change of micelles under planar confinement. 
The controlling parameters are the separation between the two hard surfaces, $D$, and the selectivities of the two surfaces, $\Lambda_u$ and $\Lambda_l$. 
We first present the monomer density profile at different values of $D$ for neutral surfaces ($\Lambda_u=\Lambda_l=0$). 
Then we use the radius of gyration tensor to quantify the shape variation under confinement.
Finally, we study the effect of varying the surface selectivity, for both the symmetric ($\Lambda_u=\Lambda_l$) and asymmetric ($\Lambda_u \ne \Lambda_l$) cases.
We employ an expansion of spherical harmonics to characterize the shape variation for asymmetric surfaces.

\subsection{Density profiles}
\label{sec:density}

\begin{figure}[htp]
  \includegraphics[width=1.0\columnwidth]{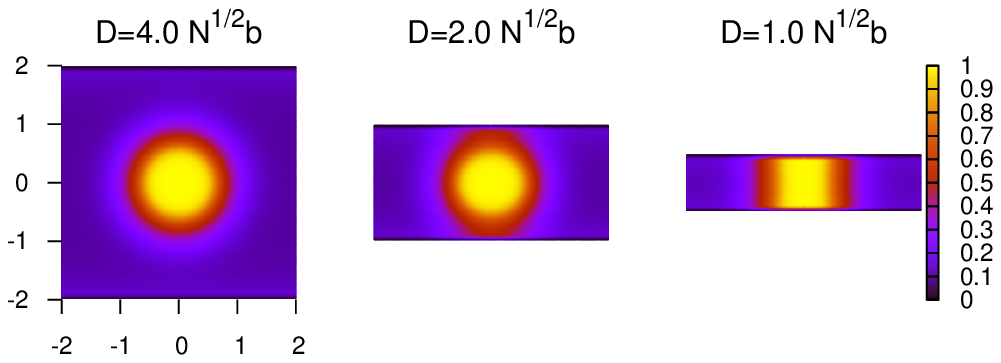}
  \includegraphics[width=1.0\columnwidth]{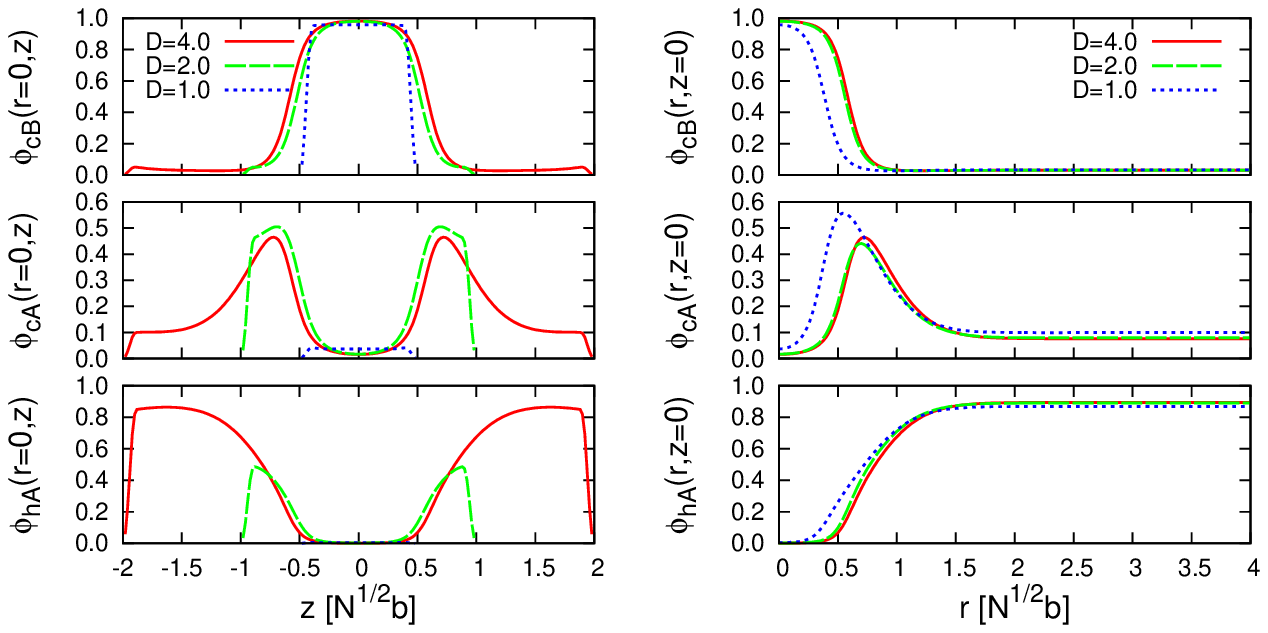}
  \caption{Top: Copolymer density profiles $\phi_c(r,z)$ for a spherical micelle with different surface separations. Bottom left panel: Cut through the density profiles at $r=0$. Bottom right panel: Cut through the density profiles at $z=0$. $\phi_{hA}$, $\phi_{cA}$, and $\phi_{cB}$ are the densities for the A-homopolymer, the A-block, and B-block of the copolymer, respectively. The A-monomer fraction of the copolymer is $f_A=0.7$ and the averaged copolymer concentration is $\phi_c^{\rm bulk}=0.11$. Three values of the surface separation are shown here: $D=4.0, 2.0$ and $1.0 \sqrt{N}b$. Note that the micelle becomes a cylinder perpendicular to the surface when $D=1.0\sqrt{N}b$. }
  \label{fig:profile_sph}
\end{figure}

We first consider a spherical micelle formed in a blend of copolymer concentration $\phi_c^{\rm bulk}=0.11$ and A-monomer fraction $f_A=0.7$. 
Figure~\ref{fig:profile_sph} shows the density profiles for a
single micelle with three surface separations $D=4.0, 2.0$, and $1.0 \sqrt{N}b$.
The surfaces are neutral ($\Lambda_l = \Lambda_u =0$).
The top panel presents the two-dimensional density profile in the coordinates of $Orz$ [see Fig.~\ref{fig:coordinates}(a)].  
The left panel shows the density profile along the line passing through the micelle center and perpendicular to the surface ($r=0$), while the right panel demonstrates the cross-section at $z=0$. 
Three density profiles: B-monomer of the copolymer $\phi_{cB}(\mathbf{r})$, A-monomer of the copolymer $\phi_{cA}(\mathbf{r})$, and the homopolymer $\phi_{hA}(\mathbf{r})$ are shown from the top to bottom. 

When the two confining surfaces are far away ($D=4.0 \sqrt{N}b$), the influence of the confinement is small as the micelle appears spherical. 
The unperturbed diameter of the spherical micelle is about $1.0 \sqrt{N}b$ (see next section for the definition).
The micelle consists of a core of the B-blocks of the copolymer, surrounded by a corona of A-blocks. 
The micelle formation is the result of two competing mechanism. 
On one hand, the copolymer chain tends to explore the free space in order to maximize the entropy. On the other hand, the localization of the copolymer results in the separation between the B-blocks and A-homopolymers, which reduces the unfavorable A-B interactions as the A-blocks acting as a shield. 
When the separation between surfaces reduces to $2.0 \sqrt{N}b$, confinement has an important effect on the micelle shape. 
The A-homopolymer, which are used to occupy the space between the micelle and surfaces, are driven away by the strong confinement [see $\phi_{hA}(r=0,z)$ plot in Fig.~\ref{fig:profile_sph}]. 
The hard surfaces are in touch with the B-block corona, pushing the copolymer towards the middle-plane of the film. 
When the surface separation becomes even smaller ($D=1.0\sqrt{N}b$), the micelle changes its shape from a sphere to a cylinder perpendicular to the surfaces.  

\begin{figure}[htp]
  \includegraphics[width=0.7\columnwidth]{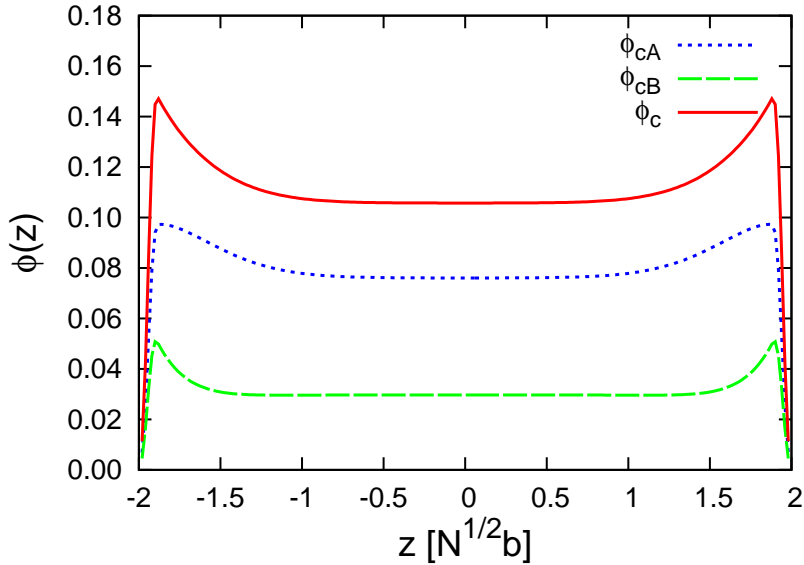}
  \caption{Copolymer density $\phi_c$ away from the micelle. Also shown are the A-monomer and B-monomer densities ($\phi_{cA}+\phi_{cB}=\phi_c$) of the copolymer. The parameters are the same as Fig.~\ref{fig:profile_sph}. The surface separation is $D=4.0\sqrt{N}b$.}
  \label{fig:bulk}
\end{figure}

Far away from the micelle, the monomer densities are equal to those of the reference system without the micelle. 
Due to the presence of hard surfaces, the density profiles of the reference system are not homogeneous in the $z$ direction. 
Figure~\ref{fig:bulk} shows the copolymer concentration as a function of $z$ for a system without the micelle ($f_A=0.7$ and $\phi_c^{\rm bulk}=0.11$).   
Since the total density is less than $1$ near the surface, this effectively reduces the A-B repulsive interaction in the surface layer \cite{Matsen1995a}. 
Therefore, it is energetically favorable for copolymers to accumulate close to the surfaces.  
Furthermore, the B-block has a stronger tendency to swell the surface layer than the A-block. 
This is because the surfaces are neutral, so there are no differences between the A-monomer a and B-monomer for the surface. 
From the entropy point of view, both ends of the copolymer have the equal probability to be found near the surface.
This corresponds to an enrichment of the minority-component of the copolymer (B-block for $f_A=0.7$).

Next we examine the cylindrical micelle parallel to the surface. 
The computation is performed in a Cartesian coordinates $Oxz$ [see Fig.~\ref{fig:coordinates}(b)].
The blend has a bulk copolymer concentration of $\phi_c^{\rm bulk}=0.016$ and A-monomer fraction $f_A=0.6$.  
At the weak confinement region, a cylindrical micelle is stable, with a slight larger diameter of $1.2\sqrt{N}b$ than the spherical micelle.
When the surface separation becomes comparable to the micelle size, the micelle is compressed and eventually changes morphology to a bilayer structure whose plane is perpendicular to the surfaces.

\begin{figure}[htp]
  \includegraphics[width=1.0\columnwidth]{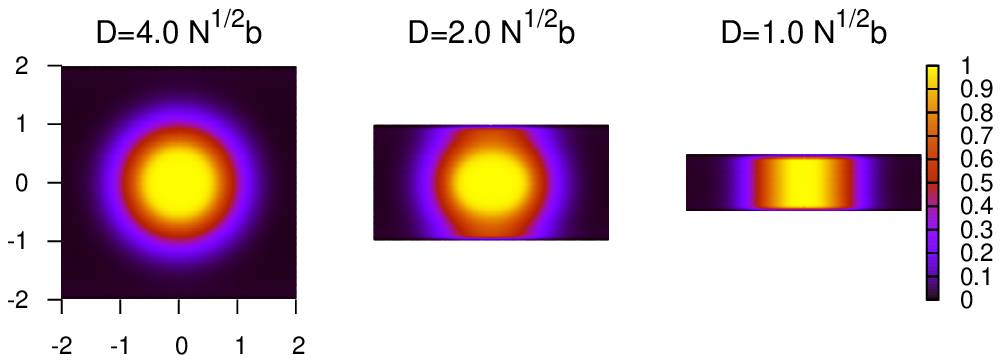}
  \includegraphics[width=1.0\columnwidth]{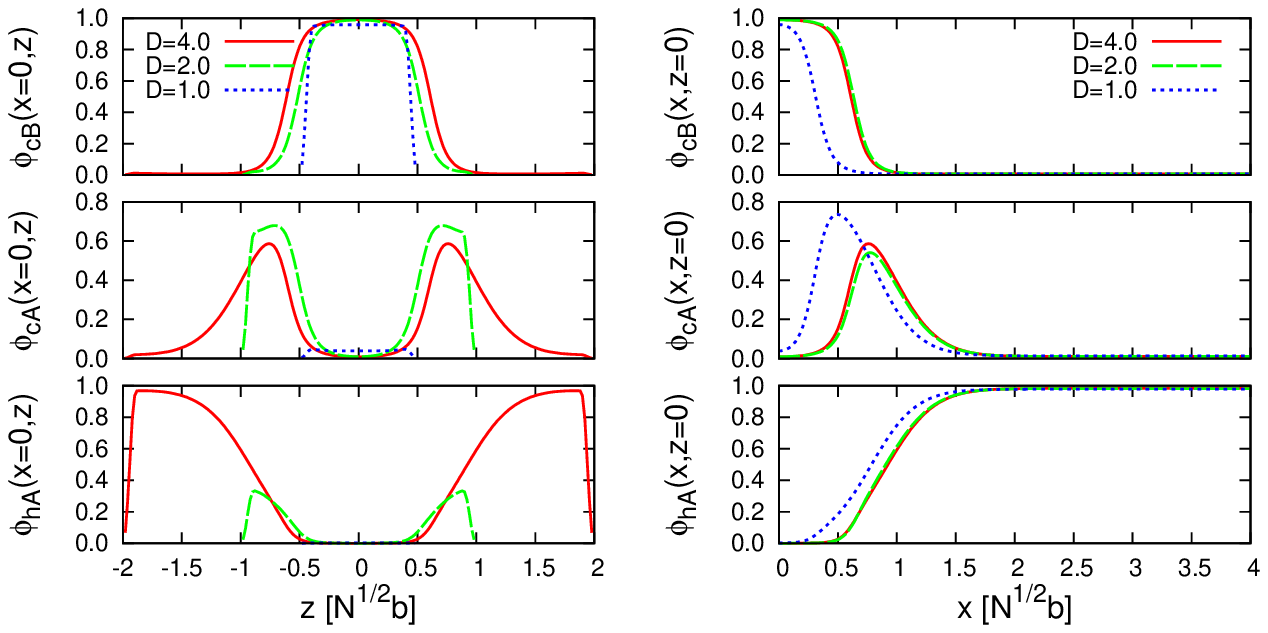}
  \caption{Top: Copolymer density profiles $\phi_c(x,z)$ for a cylindrical micelle with different surface separations. Bottom left panel: Cut through the density profiles at $x=0$. Bottom right panel: Cut through the density profiles at $z=0$. $\phi_{hA}$, $\phi_{cA}$, and $\phi_{cB}$ are the densities for the A-homopolymer, the A-block, and B-block of the copolymer, respectively. The A-monomer fraction of the copolymer is $f_A=0.6$ and the averaged copolymer concentration is $\phi_c^{\rm bulk}=0.016$. Three values of the surface separation are shown here: $D=4.0, 2.0$ and $1.0 \sqrt{N}b$. Note that the micelle becomes lamellar when $D=1.0 \sqrt{N}b$. }
  \label{fig:profile_cyl}
\end{figure}

\begin{figure}[htp]
  \centering
  \includegraphics[width=0.7\columnwidth]{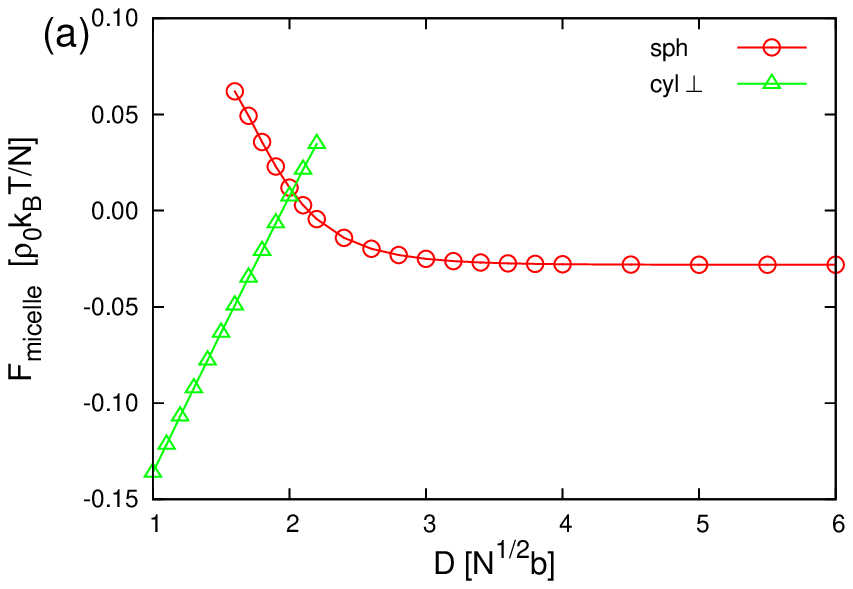}
  \includegraphics[width=0.7\columnwidth]{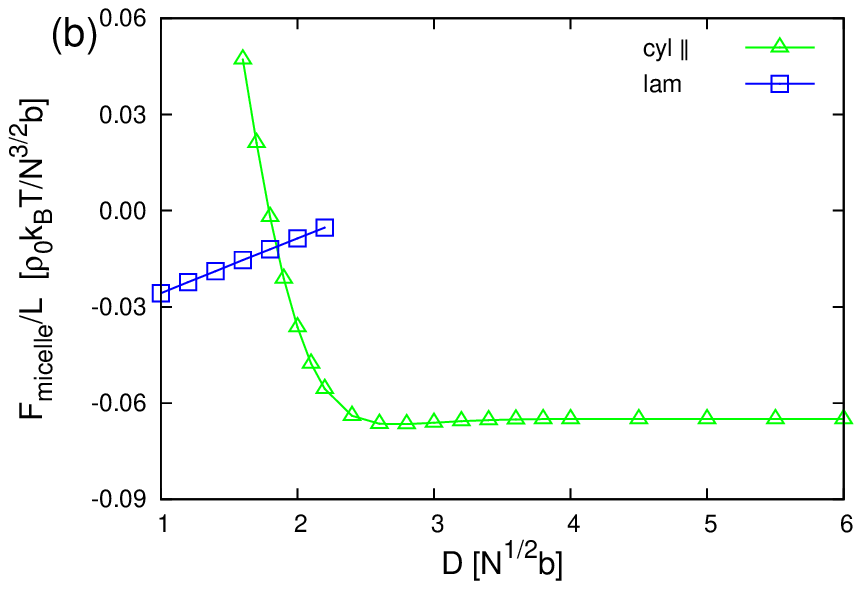}
  \caption{Free energy of an isolated micelle as a function of the surface separation. (a) The comparison between a spherical micelle and a cylindrical micelle perpendicular to the surface. (b) The comparison between a cylindrical micelle parallel to the surface and a lamellar micelle perpendicular to the surface. Note the $y$-axis for (b) is the free energy per unit length.}
  \label{fig:fe}
\end{figure}

To better understand the change of the micelle morphology, we examine the free energy of the micelle, Eq.~(\ref{eq:freeE_micelle}). 
Figure~\ref{fig:fe}(a) shows the comparison of the free energy of two micelle shapes; one is a sphere and the other one is a cylinder perpendicular to the surface. 
When the surface separation is large, the situation assembles the bulk case as the free energy of a spherical micelle is independent to the surface separation. 
When $D$ reduces to about $3.0 \sqrt{N}b$, the confinement causes the free energy to increase, and the micellar state becomes less stable. 
This observation is in agreement with the result of previous study that the critical micelle concentration increases under confinement \cite{Zhu1999, ZhangXianren2007}. 
On the other hand, the free energy of a cylindrical micelle is a linear function of the surface separation.
The linear dependency is the result of the cylinder orientation. 
The free energy of a cylindrical micelle is proportional to its length, and the micelle length is equal to the surface separation for cylindrical micelles perpendicular to the surface.
At about $D=2.0 \sqrt{N}b$, the free energy lines for the two morphologies cross each other, indicating a transition from the spherical shape to cylindrical. 

Similar phenomena are observed for the transition from a cylindrical micelle to a lamellar micelle. 
Note the orientation of the cylindrical micelle is different to that in the previous discussion. 
Figure~\ref{fig:fe}(b) shows the free energy per unit length for the two morphologies. 
A shape transition happens at around $D=1.8\sqrt{N}b$.

\subsection{Geometry frustration} 

To characterize the shape variation of the micelle under confinement, we calculate the eigenvalues for the radius of gyration tensor \cite{Theodorou1985}. 
The eigenvalues can be calculated from the monomer density profiles 
\begin{equation}
  R_{\alpha}^2 = \frac{ \int \ud V [\phi_c(\mathbf{r})-\phi'_{c}(\mathbf{r})]
    (r_{\alpha}-r_{\alpha 0})^2 }{ \int \ud V 
    [ \phi_c(\mathbf{r})-\phi'_{c}(\mathbf{r})] }, \quad \alpha=x,y,z.
\end{equation}
where $\phi_c(\mathbf{r})$ and $\phi'_c (\mathbf{r})$ are the copolymer density for systems with and without the micelle, respectively, and $r_{\alpha 0}$ are coordinates for the micelle center. 
The removal of $\phi'_c (\mathbf{r})$ is necessary in order to exclude the contribution from surface layers and to isolate the micelle from the background. 

For the spherical micelle, we compute in a cylindrical coordinate which has rotational symmetry with respect to the $z$-axis, thus $R_x^2 = R_y^2$.  
We define the asphericity $p$ by \cite{Theodorou1985},
\begin{equation}
  p=R_x^2 - \frac{1}{2}(R_y^2+R_z^2) = \frac{1}{2}(R_x^2 - R_z^2). 
\end{equation}
This quantity $p$ is useful to identify the deviation from a spherical geometry; for a perfect sphere $p=0$, and $p>0$ when the shape deviates from the spherical symmetry. 

\begin{figure}[htp]
  \centering
  \includegraphics[width=0.7\columnwidth]{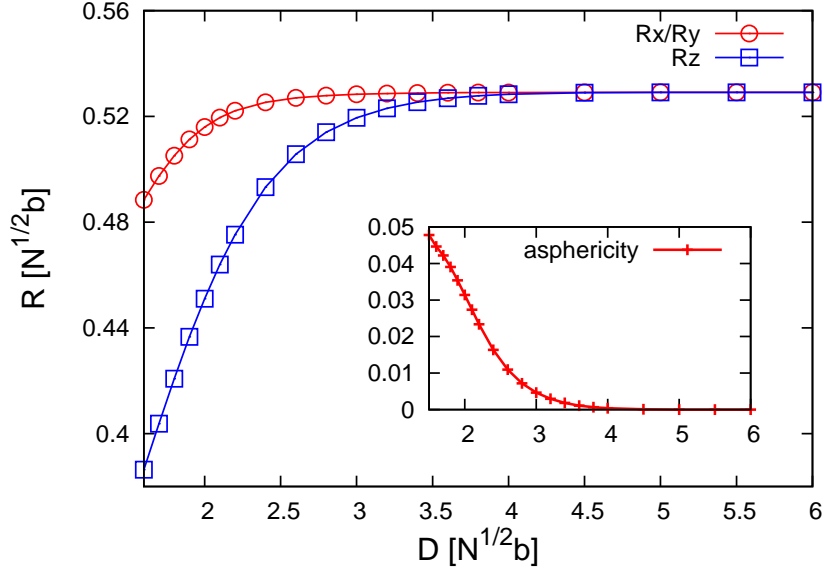}
  \caption{Gyration radii in $x$- and $z$-direction for a spherical micelle under confinement. Due to symmetry, the gyration radius $R_y$ is equal to $R_x$. Inset: the asphericity $p$ as a function of the surface separation $D$. The parameters are the same as Fig.~\ref{fig:profile_sph}.}
  \label{fig:radius_sph}
\end{figure}

The two gyration radii for a spherical micelle are shown in Fig. \ref{fig:radius_sph}.
The variation of the radii upon reducing surface separation confirms the observation stated in Section \ref{sec:density}.   
Initially, when $D$ is large, the two radii are equal.
The influence of the surfaces is negligible, and the micelle retains spherical shape.
Starting form $D=4.0\sqrt{N}b$, there are slight reduction for both radii.
Although this separation is still larger than the nature micelle size $2R_z \approx 1.0 \sqrt{N}b$, the A-homopolymer chains located between the micelle corona and the hard surfaces have less space to relax, which indirectly causes the compression of the micelle. 
Interestingly, the radius in $x$-direction is also decreased, which corresponds to a diminishing of the micelle volume. 
The result demonstrates that the confinement reduces the number of copolymer chains per micelle,  and it becomes favorable for copolymers to stay as isolated chains instead of forming micelles. 
When the film thickness is reduced further, the surface layer starts to touch the corona of the micelle, resulting in a great reduction of A-homopolymers in between the micelle and surfaces.
The surface separation $D$ becomes the characteristic length scale for the micelle in $z$-direction, as the $R_z$ shows a linear dependency on $D$.   
Eventually, the spherical micelle becomes unstable and undergoes a transition to a cylinder whose axis is normal to the surface. 
The inset of Fig.~\ref{fig:radius_sph} shows the asphericity $p$ as a function of the surface separation. 
As expected, deviation from a perfect sphere starts around $D=4.0\sqrt{N}b$ and the asphericity increases as the separation becomes smaller. 

For the cylindrical micelle parallel to the surface, the radius of gyration in the direction of the cylinder axis is not defined, thus there are only two radii $R_x$ and $R_z$. 
A quantity named acylindricity $c$ can be used to characterize the cylindrical shape \cite{Theodorou1985},
\begin{equation}
  \label{eq:acylindricity}
  c = R_x^2 - R_z^2. 
\end{equation}
The acylindricity $c=0$ for an object with perfect cylindrical symmetry. 

\begin{figure}[htp]
  \centering
  \includegraphics[width=0.7\columnwidth]{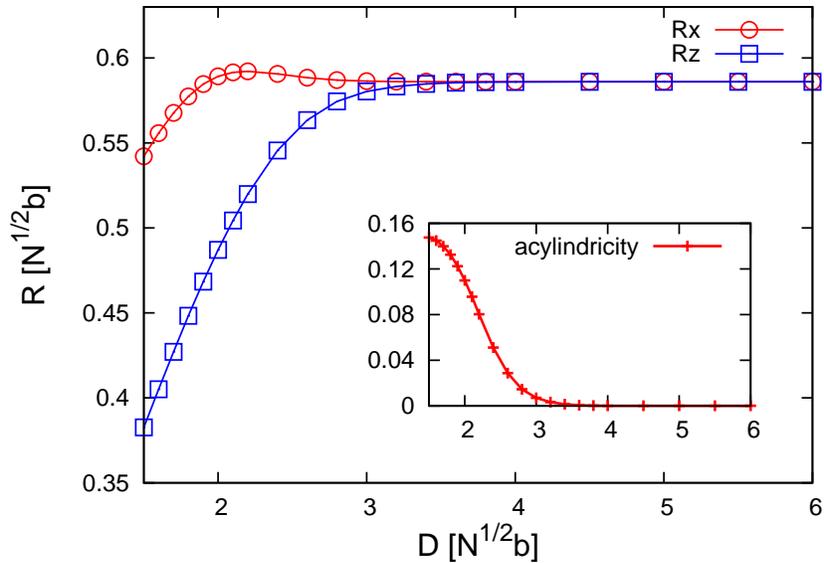}
  \caption{Gyration radii in $x$- and $z$-direction for a cylindrical micelle under confinement. Inset: the acylindricity $c$ as a function of the surface separation $D$. The parameters are the same as Fig.~\ref{fig:profile_cyl}.}
  \label{fig:radius_cyl}
\end{figure}

The results of an isolated cylindrical micelle are shown in Fig.~\ref{fig:radius_cyl}.
The trend of the radius change seems to be similar to the spherical micelle. 
Only notable difference is the small bump in $R_x$ at around $D=2.0\sqrt{N}b$. 
When the surface separation becomes smaller, the micelle first swells slightly in the $x$-direction, then proceeds to decrease. 
The inset of Fig.~\ref{fig:radius_cyl} shows the acylindricity $c$ as a function of the surface separation. 

\subsection{Surface selectivity}

\begin{figure}[htp]
  \includegraphics[width=0.7\columnwidth]{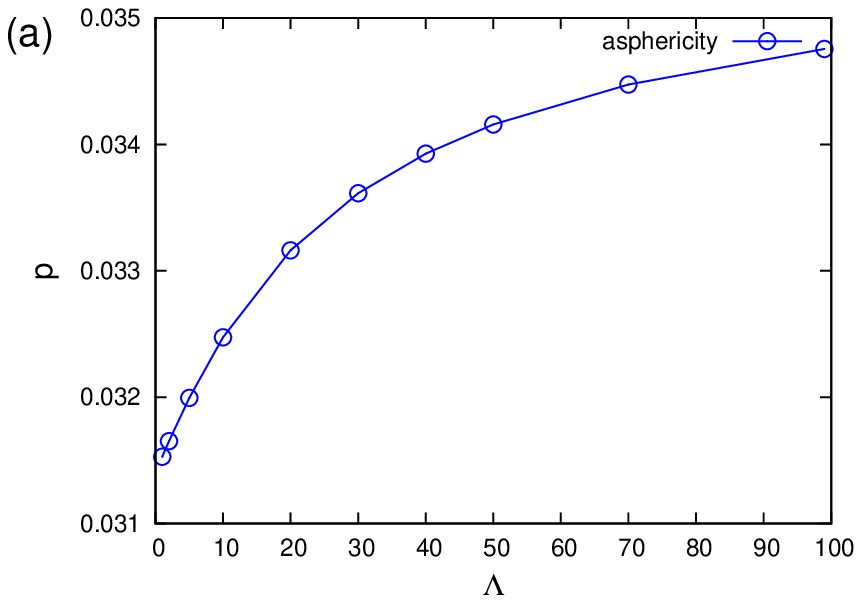}
  \includegraphics[width=0.7\columnwidth]{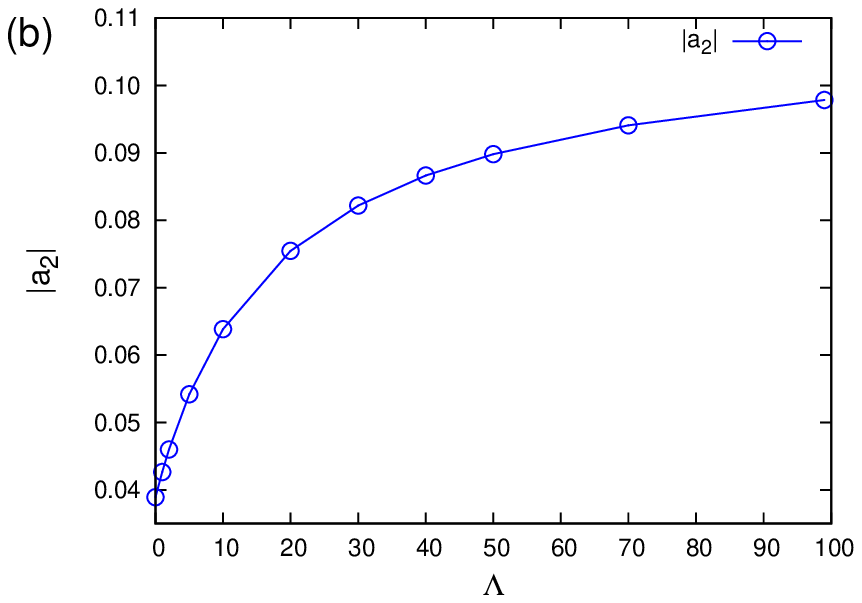}
  \caption{Effect of the surface selectivity on the micelle shape. The two surfaces are separated by $D=2.0\sqrt{N}b$, and other parameters are the same as Fig.~\ref{fig:profile_sph}. The upper and lower surfaces have the same value of selectivity $\Lambda_u=\Lambda_l=\Lambda$. (a) The asphericity $p$ as a function of $\Lambda$. (b) The second-order coefficient $a_2$ in a spherical harmonics expansion of the micelle surface.}
  \label{fig:lambda}
\end{figure}

The shape of a micelle can also be tuned by surface interactions. 
In this section, we focus on the spherical case because the cylindrical micelle shows similar behavior.
We first consider the symmetric case where the upper and lower surfaces have the same selectivity $\Lambda_l = \Lambda_u = \Lambda$. 
Since a positive value of $\Lambda$ corresponds to an attractive interaction to the A-monomer, so equivalently a repulsive interaction to the B-monomer. 
In Fig.~\ref{fig:lambda}(a), the asphericity is plotted as a function of the surface selectivity. 
The micelle becomes more aspherical as the repulsive interaction to B-monomer becomes stronger.
The effect of the surface selectivity is relatively weak in comparison to that of confinement (see the $y$-axis scales in Fig.~\ref{fig:lambda}(a) and Fig. \ref{fig:radius_sph} inset for comparison).

Asphericity allows one to quantitatively describe the degree of the shape variation, but its usage is limited, especially for asymmetric surfaces.
Another method to characterize a sphere-like shape is to expand its surface radial function $R(\theta, \phi)$ in terms of spherical harmonics $Y_{lm}(\theta,\phi)$, as follows \cite{Manyuhina2010},
\begin{equation}
  R(\theta,\phi) = \sum_{l=0}^{\infty} \sum_{m=-l}^{l} a_{lm} Y_{lm}(\theta,\phi).
\end{equation}
For the shape of rotational symmetry, the dependence on azimuth angle $\phi$ vanishes, so we only need to consider spherical harmonics with $m=0$,
\begin{equation}
  R(\theta) = \sum_{l=0}^{\infty} a_l Y_{l0}(\theta),
\end{equation}
where the coefficient $a_l$ of the above expansion is defined as 
\begin{equation}
  \label{eq:al}
  a_l = 2\pi \int_0^{\pi} \mathrm{d} \theta \sin \theta \, R(\theta) \, Y^*_{l0}(\theta). 
\end{equation}

To validate the method of spherical harmonics expansion, we reexamine the symmetric case. 
First, we locate those points where the copolymer concentration is at the middle point of the maximum and minimum, and we define the micelle surface $R(\theta)$ as a set of those points.
Then the coefficients $a_l$ are calculated using Eq.~(\ref{eq:al}).  
Since the micelle shape only deviates from a sphere slightly, we keep only the first three terms in the expansion and neglect the higher-order contributions. 
The zeroth-order coefficient $a_0$ is just the average radius. 
The first-order coefficient $a_1$ is odd with respect to $\theta=\pi/2$ ($z=0$ plane); it must vanish if the shape has a top-down symmetry.
The second-order coefficient $a_2$ is the natural choice for the symmetric case, as a large $a_2$ value corresponds to a more elongated ellipsoidal shape.
Figure~\ref{fig:lambda}(b) shows the magnitude of $a_2$ as a function of the surface selectivity $\Lambda$. 
Compare the two figures in Fig.~\ref{fig:lambda}, we see that the second-order coefficient $a_2$ exhibits a similar trend as the asphericity $p$. 

\begin{figure}[htp]
  \includegraphics[width=0.7\columnwidth]{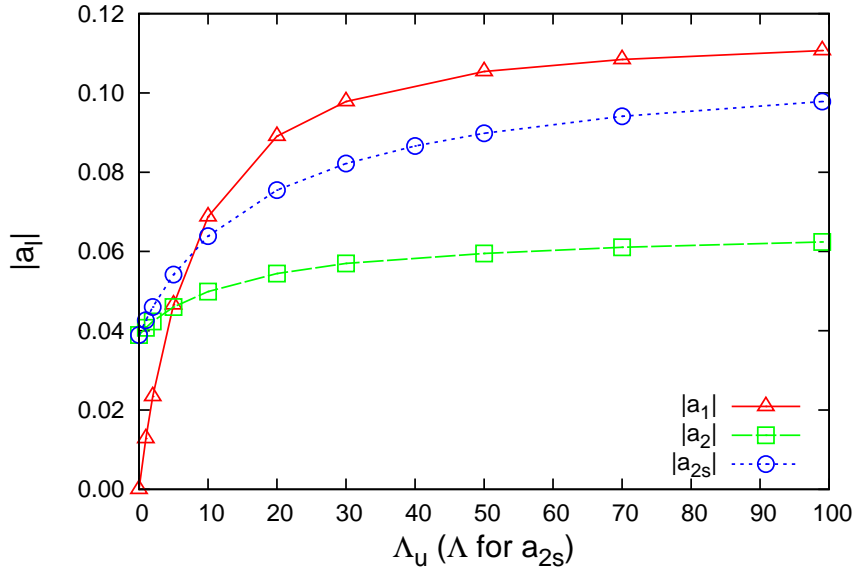}
  \caption{Effect of the surface selectivity on the micelle shape, where the two surfaces have different selectivity ($\Lambda_l \ne \Lambda_u$). The two surfaces are separated by $D=2.0\sqrt{N}b$, and other parameters are the same as Fig.~\ref{fig:profile_sph}. The lower surface is neutral $\Lambda_l=0$ and the selectivity of upper surface $\Lambda_u$ is varied. The coefficients in a spherical harmonics expansion are shown as a function of $\Lambda_u$. For comparison, the second-order coefficient for the symmetric case, $a_{2s}$, is also shown in dotted line.} 
  \label{fig:lambda_a}
\end{figure}

Next we consider the case where the upper surface has a different selectivity to the lower surface. 
We set the lower surface to be neutral $\Lambda_l=0$, while vary the upper surface selectivity $\Lambda_u$. 
Figure~\ref{fig:lambda_a} shows the first- and second-order coefficients as a function of $\Lambda_u$. 
Due to symmetry, the first-order coefficient is zero when $\Lambda_u=\Lambda_l=0$. 
The top-down symmetry is lost when the upper surface becomes repulsive to the B-monomers while the lower surface stays neutral.
At the same time, the second-order coefficient also shows a slight increase, although the magnitude is less than the symmetric case (also shown in dotted line in Fig. \ref{fig:lambda_a}). 
 
\section{Summary}
\label{sec:summary}

In this paper, we have investigated the shape variation of an isolated micelle formed in thin films of AB-diblock copolymer and A-homopolymer blends. 
SCFT has been employed to compute the free energy and density profiles of the micelle.
The shape of the micelle becomes anisotropic under planar confinement.
Deviation from the bulk geometry appears when the film thickness is comparable to the size of the micelle.   
Morphology transition can also happen when the confinement is strong. 
Two sequences of shape change are identified by comparing the free energy of micelles with different morphology. 
One is the transition from a spherical micelle to a cylindrical micelle perpendicular to the surface, and the other one is from a parallel cylindrical micelle to a lamellar micelle. 
For spherical micelles, the free energy increases as the confinement become strong; thus it becomes energetically unfavorable for the micelle formation. 
This is in agreement with Ref.~\cite{Zhu1999} that the critical micelle concentration increases as confinement increases.

Several parameters control the micelle shape in thin polymer films. 
In the bulk system, the micellization depends primarily on the copolymer architecture and monomer-monomer interactions. 
In the thin film, the confinement introduces the effect of structure frustration and surface interaction. 
The degree of structure frustration depends on the relationship between the micelle size and film thickness. 
If the two are incomparable, the aggregate must deviate from its equilibrium shape to relieve the imposed frustration. 
This in general leads to reduced size of the micelle and an ellipsoidal shape with its long axis parallel to the surfaces. 
Beside the geometry confinement, the other important factor is the surface selectivity.  
The interaction between monomers and hard surfaces can be neutral or preferential.  
The selectivity also causes deviation from the structures observed in the bulk. 
To characterize the shape variation quantitatively, we calculate the radius of gyration tensor for the micelle. 
An increase of the asphericity and acylindricity when the surface separation is reduced and when the surface interaction becomes stronger. 
For the case when the two surfaces have different selectivities, an expansion method is used to study the micelle shape, and the first-order coefficient is an indicator of the deviation from top-down symmetry. 

Our SCFT calculations demonstrate that valuable information can be obtained, such as the geometrical dimensions of the micelle, and the monomer density profiles. 
These properties are closely related to several parameters: the architecture of the copolymer, the monomer-monomer interactions, the film thickness, and surface interactions. 
In this paper, we focus on the effect of the film thickness and surface interactions, and use diblock copolymer as an example. 
Our model can be easily extended to different polymer architectures, and the shape variation by varying the surface separation and selectivities would exhibit similar trends for different monomer-monomer interactions. 
By varying these parameters, we can control the micelle shape and size in the thin film. 
Since it is expensive and time-consuming for experimentalists to explore the parameter space by trial-and-error, theoretical study can be of great help to understand the underlying physics and provide guidance for experiments.

The present approach has a number of limitations.
For the cylindrical micelle parallel to the surface and the lamellar micelle, one has to be cautious because they are ideal cases without the ends and edges. 
In real experiments, they can be twisted and forms labyrinth structures. 
Our method also neglects the translational entropy of the micelle; therefore the calculated points of morphology transition are only qualitatively correct. 
Furthermore, the surface attraction to one block of the copolymer may induce the formation of surface micelles, an important phenomenon we have not touched in this paper.  
Nevertheless, by computing the free energies and shape parameters for different micelle morphologies, we have obtained the trends of micelle formation in the thin polymer films. 
The results provide a detailed picture of the variation and transition of micelle shape and illustrate the effect of confinement and surface interactions.

\begin{acknowledgments}
We thank Arif Gozen, Kristen Roskov, Jan Genzer, and Richard Spontak from NCSU for their experiments on polymer film \cite{2011_tdgl} which inspired this work. 
This work was supported by the Natural Sciences and Engineering Research Council (NSERC) of Canada. The computation was made possible by the facilities of the Shared Hierarchical Academic Research Computing Network (SHARCNET:www.sharcnet.ca) and Compute/Calcul Canada.  
\end{acknowledgments}


\end{document}